\documentclass[12pt]{article}
\usepackage{amssymb}
\usepackage{amsmath}
\usepackage{graphicx,color}

\usepackage{graphics}
\usepackage{color}
\usepackage{amssymb,amsmath}
\usepackage{slashed}
\usepackage{epstopdf}
\usepackage{booktabs}
\usepackage{mathrsfs}
\usepackage[export]{adjustbox}
\usepackage{epsfig}
\usepackage{graphicx,color}

\newcommand{\bea}{\begin{eqnarray}}
\newcommand{\eea}{\end{eqnarray}}
\newcommand{\sy}{s(y)}

\numberwithin{equation}{section}

\begin{document}
%
%\vspace*{1.0cm}
\begin{titlepage}
%%%%%%%%%%%%%%%%%
%\begin{flushright}
% preprint number 
%\end{flushright}
%%%%%%%%%%%%%%%%%
\vspace*{10mm}
\begin{center}
\baselineskip 25pt 
{\Large\bf
%%%%%%%%%%%%%%%%%%%%%%%%%%%
Aspects of Domain-Wall Standard Model
%%%%%%%%%%%%%%%%%%%%%%%%%%%
}
\end{center}
\vspace{5mm}
\begin{center}
{\large 
Nobuchika Okada$^a$\footnote{okadan@ua.edu},  
Digesh Raut$^b$\footnote{drautphys@gmail.com}, and 
Desmond Villalba$^c\footnote{dvillalb@umw.edu}$
}

%\end{center}

\vspace{.5cm}

{\baselineskip 20pt \it
$^a$Department of Physics and Astronomy, \\
The University of Alabama, Tuscaloosa, AL 35487, USA\\
} 
{\baselineskip 20pt \it
$^b$Department of Physics, \\
St. Mary's College of Maryland, St Marys City , MD 20686, USA\\
}
{\baselineskip 20pt \it
$^c$Department of Chemistry and Physics, \\
University of Mary Washington, Fredericksburg , VA 22401, USA\\
}

\end{center}
\vspace{0.5cm}

%%%%%%%%%%%%%%%%%%%%%%
\begin{abstract}
%%%%%%%%%%%%%%%%%%%%%%

In the ``Domain-Wall Standard Model,'' 
  all the Standard Model (SM) fields are localized in certain domains 
  in a non-compact 5-dimensional space-time. 
While the SM is realized as a 4-dimensional effective theory at low energies, 
  the model involves Kaluza-Klein (KK) modes of the SM particles. 
In this paper we introduce two simple solvable examples which lead to domain-wall configurations 
  for the SM particles and their KK-modes. 
Based on the examples, we address a variety of phenomenologies of the Domain-Wall SM, 
  such as the KK-mode gauge boson phenomenology at the Large Hadron Collider (LHC), 
  the effect of the KK-mode SM fermions on Higgs boson phenomenology, 
  and the KK-mode fermion search at the LHC 
  with its decay into a corresponding SM fermion and a Nambu-Goldstone boson 
  associated with a spontaneous breaking of the translational invariance in the 5th dimension.  
We also propose a simple unified picture of localizing all the SM fields. 

\end{abstract}
\end{titlepage}

%%%%%%%%%%%%%%%%%%%%%%%%%%%%%%%%%%%%
\section{Introduction}
%%%%%%%%%%%%%%%%%%%%%%%%%%%%%%%%%%%%

An idea that our world consists of more than 4D space-time has always been fascinating us. 
After the discovery of the D-brane in string theories \cite{D-brane}, the brane-world scenario 
  has been intensively studied as new physics beyond the Standard Model (SM). 
In extra-dimensional models, a new property, ``geometry,'' comes into play in phenomenology 
  and provides us with a new possibility of understanding mysteries in the SM. 
One well-known brane-world scenario is the so-called large extra-dimension model \cite{ADD}, 
  which offers a solution to the gauge hierarchy problem with the extra-dimensional Planck mass 
  being at the TeV scale while reproducing the 4D Planck mass by a large extra-dimensional volume.   
Another well-known scenario is the warped extra-dimension model \cite{RS} in 5D, 
  where the Planck scale is warped down to the TeV scale 
  in the presence of the anti-de Sitter (AdS) curvature in the 5th dimension.

In usual extra-dimensional models, extra-dimensions are compactified on some manifolds or orbifolds, 
   and our 3-spatial dimensions and  extra-dimensions are treated differently. 
We may think it more natural if all spatial dimensions are non-compact 
  and the SM  is realized as a 4D effective theory. 
This picture requires that all the SM fields as well as 4D graviton
  are localized in certain 3-spatial dimensional domains in the bulk space. 
 The so-called RS-2 scenario proposed in Ref.~\cite{RS2} provides 
  a simple realization of this picture for 4D graviton. 
In the scenario, due to the 5D AdS curvature, 4D graviton is localized 
  around a point in the non-compact 5th dimension and the 4D Einstein gravity 
  is reproduced at low energies.

Recently, the authors of the present paper have proposed a framework to construct 
 ``Domain-Wall Standard Model" in a non-compact 5D space-time \cite{DWSM}, 
  where all the SM fields are localized in certain domains of 5th dimension. 
This model is a field-theoretical realization of a ``3-brane" on which all the SM fields are confined. 
However, the finite widths of the ``3-branes" corresponding to the SM fields 
  lead to rich particle physics phenomenologies. 
Based on a simple setup for localizing the SM fields, we have obtained the SM as a 4D effective theory. 
The localization mechanism predicts the Kaluza-Klein (KK) modes of the SM particles, 
  and we have investigated the phenomenology of KK-mode gauge boson 
  at the Large Hadron Collider (LHC) experiment.

In this paper, we investigate various aspects of the Domain-Wall SM in detail.  
For concreteness, we introduce two solvable examples to localize all the SM fields 
  in certain domains of the 5th dimension, and provide the explicit forms 
  of the KK-mode mass spectrum and eigenfunctions.   
We then derive the 4D effective Lagrangian involving the KK-mode SM particles. 
Among a variety of possible phenomenologies of the Domain-Wall SM, 
  we address the LHC phenomenology of the KK-mode gauge bosons,
  the Higgs boson phenomenology in the presence of  the KK-mode SM fermions, 
  and the KK-mode fermion search at the LHC.

We begin with the localization of the gauge field in the next section,  
   where the basic formalism to maintain the 4D gauge invariance is presented. 
For detailed analysis, we introduce two solvable examples and obtain the explicit form 
   for the KK-mode mass spectrum and the KK-mode eigenfunctions. 
In Sec.~\ref{sec:3}, we consider the Higgs sector and apply the same procedure 
   taken for the gauge field to localize the Higgs boson and its vacuum expectation value (VEV).  
We derive the mass spectrum for the zero-mode and KK-modes of the gauge boson 
   after the gauge symmetry breaking.   
In Sec.~\ref{sec:4}, we consider the localization mechanism of the SM chiral fermions 
  and derive the SM Lagrangian in the 4D effective theory, 
  which also involves the KK-mode gauge bosons and Higgs bosons. 
We investigate a variety of phenomenologies for the Domain-Wall SM in Sec.~\ref{sec:5}. 
The last section is devoted to conclusions and discussions.

%%%%%%%%%%%%%%%%%%%%%%%%%%%%%%%%%%%%%%
\section{The gauge sector}
%%%%%%%%%%%%%%%%%%%%%%%%%%%%%%%%%%%%%%
\label{sec:2}

Let us first consider the gauge sector of the Domain-Wall SM, 
 where a gauge field is localized around a point in the 5th dimension 
 while keeping the 4D gauge invariance. 
Since the essential mechanism for localizing a gauge field is independent of the gauge group structure, 
  we address the gauge field localization based on a U(1) gauge theory. 
For localizing the gauge field, we adopt a simple way proposed in Ref.~\cite{OS}\footnote{
  We can find that the same idea was discussed elsewhere before Ref.~\cite{OS}. 
  See, for example, Ref.~\cite{LO}.      
}    
 and introduce the following Lagrangian for the U(1) gauge field in 5D flat Minkowski space: 
\bea
\mathcal{L}_{5}=-\frac{1}{4}s(y)F_{MN}F^{MN}, 
\eea
where $F_{MN}$ is the gauge field strength, and $M, N=0,1,2,3,y$ with $y$ being the index for the 5th coordinate. 
Our convention for the metric is $\eta_{MN}={\rm diag}(1,-1,-1,-1,-1)$, and in general we suppress coordinate dependence 
  of the fields unless emphasis is needed. 
In the original Lagrangian, we identify $s(y)=1/{\bar g}^2$ with $\bar{g}$ being the 5D gauge coupling, 
  and this $y$-dependence is the key to localize the gauge field. 
In the defined Lagrangian, the gauge field and $s(y)$ have mass dimension one.

Decomposing the field strength into its components yields the following expression (up to total derivative terms):
\bea
\mathcal{L}_{5}
 &=&
 \frac{1}{2} s A^{\mu}\left(g_{\mu \nu}\Box_{4} -\partial_{\mu}\partial_{\nu}\right)A^{\nu}-\frac{1}{2} s A_{y}\Box_{4}A_{y} 
\nonumber \\
 &-& \frac{1}{2}A_{\mu}\partial_{y}\left(s \partial_{y}A^{\mu}\right)-\left(\partial_{\mu}A^{\mu}\right)\partial_{y}\left(s A_{y}\right),
\label{L5} 
\eea
where $A_{\mu}$ $(\mu=0,1,2,3)$ and $A_y$ are a gauge field and a scalar field 
  in 4D space-time, and the first line in the right-hand-side denotes the 4-dimensional kinetic terms for these fields.
Since the original Lagrangian formally maintains the 5D gauge invariance, we may use the gauge degrees of freedom 
  to eliminate $2$ components from $A_M$. 
Although the so-called ``axial gauge'' ($A_y=0$) is sometimes employed in literature, 
  we treat $A_y$ as a dynamical field and eliminate $2$ degrees of freedom from $A_\mu$ as we usually do in 4D gauge field theories. 
As we will show in the following, the zero-mode of $A_y$ vanishes because of the breakdown of the gauge invariance 
  due to 5th coordinate-dependence of the gauge coupling $s(y)$. 
Hence, the gauge choice of $A_y=0$ may make the gauge structure of the theory unclear. 
The last term in Eq.~(\ref{L5}) contains a mixing between $A_{\mu}$ and $A_y$. 
Note that this structure is analogous to that in spontaneously broken gauge theories, 
  and suggests us to eliminate it by adding a gauge fixing term, 
  which is a 5D analog to the $R_{\xi}$ gauge \cite{DWSM, Arai}: 
\bea
\mathcal{L}_{\rm GF}=-\frac{s}{2 \, \xi}\left(\partial_{\mu}A^{\mu}- \frac{\xi}{s} \partial_{y}(s A_y)\right)^2 ,
\eea
where $\xi$ is a gauge parameter. 
The total Lagrangian now reads $\mathcal{L}=\mathcal{L}_{5}+\mathcal{L}_{\rm GF}=\mathcal{L}_{\rm gauge}+\mathcal{L}_{\rm scalar}$, 
 where
\bea 
\mathcal{L}_{\rm gauge}&=&\frac{1}{2} s A^{\mu}\left(g_{\mu \nu}\Box_{4}-\left(1-\frac{1}{\xi}\right)\partial_{\mu}\partial_{\nu}\right)A^{\nu}
   -\frac{1}{2}A_{\mu}\partial_{y}(s \, \partial_{y}A^{\mu}), \nonumber \\ 
\mathcal{L}_{\rm scalar}&=&-\frac{1}{2} s  A_{y}\Box_{4} A_{y}+\frac{1}{2} s \, \xi A_{y}\partial_{y} \left(\frac{1}{s}\partial_{y}(s A_{y}) \right).
\label{decom}
\eea

Using the Kaluza-Klein (KK) mode decomposition of the gauge and scalar fields, 
\bea
A_{\mu}(x,y)=\sum _{n=0}^\infty A_{\mu}^{(n)}(x) \, \chi^{(n)}(y), \quad A_{y}(x,y)=\sum _{n=0}^\infty \eta ^{(n)}(x) \, \psi^{(n)}(y),  
\eea
we can rewrite the Lagrangians in Eq.~(\ref{decom}) as 
\bea 
\mathcal{L}_{\rm gauge}&=& \sum_{n=0}^\infty \frac{1}{2} s  \left( \chi^{(n)} \right)^2
 \left[ A_{\mu}^{(n)} \left(g^{\mu \nu}(\Box_{4} + m_n^2) -\left(1-\frac{1}{\xi}\right)\partial^{\mu}\partial^{\nu}\right)A_{\nu}^{(n)}  \right] , 
 \nonumber \\ 
\mathcal{L}_{\rm scalar}&=&- \sum_{n=0}^\infty  \frac{1}{2} s   \left( \psi^{(n)} \right)^2  
  \left[ \eta^{(n)} \left( \Box_{4} + \xi \, m_n^2 \right) \eta^{(n)} \right], 
\label{L2} 
\eea
where $\chi^{(n)}$ and $\eta^{(n)}$ are the solutions of the KK-mode equations: 
\bea
\frac{d}{dy}
\left(s \,  \frac{d}{dy}\chi^{(n)} \right)+ s \, m_{n}^{2}\chi^{(n)}=0,  \; \; \; \; 
\frac{d}{dy}\left( \frac{1}{s} \frac{d}{dy} \left(s \psi^{(n)} \right) \right)+ m_{n}^{2}\psi^{(n)}=0. 
\label{KK_EOM}
\eea
Here we have expected the KK-modes of $\eta^{(n)}$ are would-be Nambu-Goldstone (NG) mode 
  eaten by the KK-modes of $A_\mu^{(n)}$ 
  and enjoy the degrees of freedom for their longitudinal modes. 
This picture is consistent only if the two equations in Eq.~(\ref{KK_EOM}) have solutions 
   with common $m_n$ eigenvalues for a given function $s(y)$. 
In the following, we will discuss a couple of solvable examples for $s(y)$ and show this consistency explicitly. 
However, we emphasize that the pairing of the KK-mode mass spectrum
  between $\chi^{(n)}$ and $\psi^{(n)}$ are held only for massive modes 
  while the zero-mode $\psi^{(0)}$ vanishes.

Even for a general function of $s(y)$, we can easily find zero-mode solutions ($m_0=0$) for Eq.~(\ref{KK_EOM}) such that 
\bea
\chi^{(0)}=\tilde{c}_{\chi}+c_{\chi}\int^{y} \frac{dy^\prime}{s(y^\prime)}, \; \; \; 
\psi^{(0)}=\frac{\tilde{c}_{\psi}}{\sy}+\frac{c_{\psi}}{\sy}\int^{y} dy^\prime s(y^\prime),
\eea
where $ \tilde{c}_{\chi}$, $c_{\chi}$, $\tilde{c}_{\psi}$, and $c_{\psi}$ are constants. 
In order to localize the gauge field in the finite domain,  we impose $\sy \to 0$ as $|y| \to \infty$. 
In addition, the gauge and scalar fields in the 4D effective theory must be normalizable in the sense that
\bea
\int_{-\infty}^{\infty} dy \, \sy \, \chi^{(0)}(y) \, \chi^{(0)} (y) < \infty, \; \; \; \;
\int_{-\infty}^{\infty} dy \, \sy \, \eta^{(0)}(y) \, \eta^{(0)} (y) < \infty. 
\label{norm}
\eea
Considering the zero-mode solution for the gauge field, these constraints require $c_{\chi}=0$, 
   resulting in the zero-mode for the gauge boson having a constant configuration in the 5th dimension. 
Note that this unique solution leads to the universal gauge coupling, in other words, 
   4D gauge invariance in the 4D effective theory, independently of configurations of matter fields in the bulk.  
On the other hand, the solution of $\psi^{(0)}$ cannot satisfy the requirement given in Eq.~(\ref{norm}) 
   unless $c_{\psi}=\tilde{c}_{\psi}=0$, and hence $\psi^{(0)}=0$ is the only appropriate choice 
   for the zero-mode of the scalar. 
Therefore, no (normalizable) zero-mode exists for the scalar component. 
We may consider a special setup where $s(y)$ is independent of $y$. 
This is a trivial case that the 5D gauge invariance is manifest 
  and a constant $\psi^{(0)}$ is a solution of the KK-mode equation, 
  although the gauge field is not localized. 
Therefore, the absence of the zero-mode scalar originates from the explicit breaking of the 5D 
  gauge invariance due to $y$-dependence of the gauge coupling $s(y)$.

The Domain-Wall SM has rich phenomenological aspects as we will discuss below. 
For our phenomenology discussions, we need solvable system that provides us 
  with explicit forms for the gauge boson KK-mode spectrum and eigenfunctions. 
A simple example is discussed in our previous work \cite{DWSM}.     
In the following, we introduce two solvable examples, 
  which are much more non-trivial than the example in Ref.~\cite{DWSM}.

In solving the KK-mode equations in Eq.~(\ref{KK_EOM}), 
  it is convenient to rewrite the equations with a function $f(y)$ defined as $s(y)=f(y)^2$ 
  and introducing new variables, 
\bea
  \tilde{\chi}^{(n)}(y) = f(y) \, \chi^{(n)}(y), \; \; \; \;  \tilde{\psi}^{(n)}(x) = f(y) \, \psi^{(n)}(x),
\label{newKK}  
\eea  
corresponding to the gauge and the scalar fields.
We then have the KK-mode equations which have the form of the Schr\"odinger equation: 
\bea
& \left[ -\partial_y^2 - G(y)^\prime  + G(y)^2 \right] \tilde{\chi}^{(n)} = m_n^2  \tilde{\chi}^{(n)}, \nonumber\\
& \left[ -\partial_y^2 + G(y)^\prime + G(y)^2 \right] \tilde{\psi}^{(n)} = m_n^2  \tilde{\psi}^{(n)}, 
\label{KK_EOM2}
\eea
where $\prime$ denotes $d/dy$, and $G(y) =- f(y)^\prime/f(y)$.  
It is easy to find the zero-mode solution as $\tilde{\chi}^{(0)}(y) \propto f(y)$. 
This result implies that if we have a solvable 1D Quantum Mechanical system resulting in bound states, 
  we adopt this system as our solvable example by identifying $f(y)$ with the ground-state eigenfunction.

%%%%%%%%%%%%%%%%%%%%%
\subsection*{Solvable example 1}
%%%%%%%%%%%%%%%%%%%%%
We now consider the first solvable example which is a Gaussian type function, 
\bea 
  s(y) = f(y)^2 = M \exp \left[ - (m_V y)^2 \right],
\eea
where $M$ and $m_V$ are (positive) mass parameters. 
Substituting it into the KK-mode equations, we obtain 
\bea
& \left[ -\partial_y^2 + m_V^4 y^2 \right] \tilde{\chi}^{(n)} = \left(m_n^2 + m_V^2 \right)  \tilde{\chi}^{(n)}, \nonumber\\
& \left[ -\partial_y^2 + m_V^4 y^2 \right] \tilde{\psi}^{(n)} = \left(m_n^2 -m_V^2 \right) \tilde{\psi}^{(n)}, 
\label{KK_EOM3}
\eea
which are nothing but the Schr\"odinger equation for the 1D harmonic oscillator, 
\bea 
   H \Psi^{(n)} = \omega \left( a^\dagger a + \frac{1}{2} \right) \Psi^{(n)} =E_n \Psi^{(n)}, 
\eea
with the identifications of the frequency $\omega=2 m_V^2$ and the annihilation/creation operator,  
\bea
 a = \frac{1}{\sqrt{2 m_V^2}} \left( \frac{d}{d y} + m_V^2 y \right), \; \; \; 
  a^\dagger = \frac{1}{\sqrt{2 m_V^2}} \left( -\frac{d}{d y} + m_V^2 y \right).  
\eea
Thus, using the energy eigenvalues  given by $E_n=2 m_V^2 \left( n+ \frac{1}{2} \right)$ ($n=0, 1, 2 , \cdots)$ 
  for the harmonic oscillator, we find the KK-mode spectra for the gauge bosons 
  and the would-be NG modes 
  as $m_n^2 = 2 \,n \, m_V^2$ and $m_n^2=2 (n+1) m_V^2$ for $n=0, 1, 2, \cdots$, respectively. 
Note that no zero-mode exists for the scalar field. 
Shifting $n\to n+1$ for the scalar mode, we can see the pairing of the massive modes of the gauge bosons and would-be NG modes   
  ($2 n m_V^2$ and $\xi\, 2 n  m_V^2$ for $n=1, 2, \cdots$). 
This is nothing but what we expected.

Using the harmonic oscillator algebra, $[a, a^\dagger]=1$, it is straightforward to obtain the KK-mode functions. 
For example, the zero-mode function $\tilde{\chi}^{(0)}$ is obtained as a solution of  $a \tilde{\chi}^{(0)}=0$, 
  and the $n$-th KK-mode function is generated as $\tilde{\chi}^{(n)} \propto (a^\dagger)^n \tilde{\chi}^{(0)}$ 
  by using the zero-mode function. 
After integrating out the 5th-dimensional degrees of freedom for the Lagrangian of Eq.~(\ref{L2}), 
  we obtain the 4D effective Lagrangians with the canonically normalized kinetic terms as 
\bea
\mathcal{L}_{\rm gauge}^4&=& \sum_{n=0}^\infty \frac{1}{2} 
  \left[\int_{-\infty}^\infty  s  \left( \chi^{(n)} \right)^2 \right]
  \left[ A_{\mu}^{(n)} \left(g^{\mu \nu}(\Box_{4} + m_n^2) -\left(1-\frac{1}{\xi}\right)\partial^{\mu}\partial^{\nu}\right)A_{\nu}^{(n)}  \right] 
 \nonumber \\ 
&=& 
 \sum_{n=0}^\infty \frac{1}{2} 
   \left[ A_{\mu}^{(n)} \left(g^{\mu \nu}(\Box_{4} + m_n^2) -\left(1-\frac{1}{\xi}\right)\partial^{\mu}\partial^{\nu}\right)A_{\nu}^{(n)}  \right] 
 \nonumber \\  
\mathcal{L}_{\rm scalar}^4&=&- \sum_{n=1}^\infty  \frac{1}{2} 
\left[\int_{-\infty}^{\infty} s   \left( \psi^{(n)} \right)^2 \right] 
  \left[ \eta^{(n)} \left( \Box_{4} + \xi \, m_n^2 \right) \eta^{(n)} \right], \nonumber \\
&=&- \sum_{n=1}^\infty  \frac{1}{2} 
  \left[ \eta^{(n)} \left( \Box_{4} + \xi \, m_n^2 \right) \eta^{(n)} \right], 
\label{L_eff} 
\eea
for the KK-mode decomposition, 
\bea
A_{\mu}(x,y)=\sum _{n=0}^\infty A_{\mu}^{(n)}(x)\chi^{(n)}(y), \quad A_{y}(x,y)=\sum _{n=1}^\infty \eta ^{(n)}(x)\chi^{(n-1)}(y).  
\eea
Here, we have changed the label $n$ for the KK-mode decomposition of $A_y$ by $n \to n+1$, 
  and $\psi^{(n)}$ is given by $\psi^{(n)}=\chi^{(n-1)}$ for $n=1, 2, \cdots$. 
The explicit form of the first three KK-mode functions are  
\bea 
 \chi^{(0)}(y) &=& g, \nonumber \\ 
  \chi^{(1)}(y) &=&  \sqrt{2} \, g \left( m_V y \right), \nonumber \\  
  \chi^{(2)}(y) &=& \frac{g}{\sqrt{2}} \left( 1- 2 (m_V y)^2 \right) ,  \nonumber \\ 
   \chi^{(3)}(y) &=& \frac{g}{\sqrt{3}} (m_V y) \left( 3- 2 (m_V y)^2 \right) ,   
\eea
where $g$ is the U(1) gauge coupling in the 4D effective theory defined by $g=\pi^{-1/4} \sqrt{m_V/M}$.

%%%%%%%%%%%%%%%%%%%%%
\subsection*{Solvable example 2}
%%%%%%%%%%%%%%%%%%%%%

As the second example, we consider 
\bea 
  s(y)= f(y)^2 = \frac{M}{\left[ \cosh(m_V y) \right]^{2 \gamma}}, 
\label{exp2}  
\eea
where $M$ and $m_V$ are positive mass parameters, and $\gamma$ is a positive constant. 
We then express Eq.~(\ref{KK_EOM2}) as 
\bea
&& \left[ -\partial_y^2 - \frac{\gamma (\gamma+1) m_V^2}{\cosh^2(m_Vy)} \right] \tilde{\chi}^{(n)} 
  = \left(m_n^2 - \gamma^2 m_V^2 \right)  \tilde{\chi}^{(n)}, \nonumber\\
&& \left[ -\partial_y^2 - \frac{\gamma (\gamma-1) m_V^2}{\cosh^2(m_Vy)} \right] \tilde{\psi}^{(n)} 
  = \left(m_n^2 - \gamma^2 m_V^2 \right)  \tilde{\psi}^{(n)}.  
\label{KK_EOM4}
\eea
These equations have the form of the Schr\"odinger equation, 
  $(-\partial_y^2 +V) \Psi^{(n)}=E_n \Psi^{(n)}$. 
Since the potential corresponds to $V \propto -1/\cosh^2(m_V y) < 0$, 
  we expect the existence of a bound state with $E_n <0$.

We are interested in the localization of the gauge field, namely, bound states 
  from the  the Schr\"odinger equation satisfying 
  the following boundary conditions: $ |\tilde{\chi}^{(n)}(0)|< \infty$ for $y \to 0$,  
  and $\tilde{\chi}^{(n)}(y) \to 0$ for $|y| \to \infty$. 
Such solutions are described by using the hyper-geometric function $F[a,b; c; y]$ \cite{HO}. 
We find the eigenvalues for $\tilde{\chi}^{(n)}$ as 
\bea 
  m_n^2 = n \left( 2 \gamma -n \right) m_V^2  \; \; \; \; (n=0, 1, 2, \cdots < \gamma).
\label{mn1}  
\eea 
The number of (localized) KK-modes is terminated by a condition $E_n = m_n^2 - \gamma^2 m_V^2 < 0$, 
  and thus  a massive mode exists for $\gamma > 1$. 
The eigenfunctions for even numbers of 
  $n= 2 n^\prime \; (n^\prime =0,1,2, \ldots )$ and odd numbers of 
  $n= 2 n^{\prime \prime}+1 \; (n^{\prime \prime} =0,1,2,\ldots )$ 
are given by (up to normalization factor) 
\begin{eqnarray} 
 \tilde{\chi}^{(n^\prime)}(y) = \left[ \cosh(m_V y) \right]^{- \gamma} \, 
   F\left[-n^\prime , -\gamma + n^\prime; 1/2 ; 1- \cosh^2(m_V y) \right], 
\end{eqnarray}
and 
\begin{eqnarray}
  \tilde{\chi}^{(n^{\prime \prime})}(y) =  \sinh(m_V y) 
      \left[\cosh(m_V y) \right]^{- \gamma}  
  F\left[-n^{\prime \prime}, - \gamma + n^{\prime \prime}+1 ; 3/2 ; 1- \cosh^2(m_V y) \right],   
\end{eqnarray}
respectively. 

Similarly, for the scalar field we impose the boundary conditions: $|\tilde{\psi}^{(n)}(0)| < \infty$ for $y \to 0$,  
  and $\tilde{\psi}^{(n)}(y) \to 0$ for $|y| \to \infty$. 
We can easily find the eigenfunctions of $\tilde{\psi}^{(n)}$ as follows. 
Substituting $\gamma= \overline{\gamma}+1$ and $m_n^2 =\overline{m_n}+ (2 \overline{\gamma}+1)m_V^2 $ 
  into the second equations, we obtain 
\bea
 \left[ -\partial_y^2 - \frac{\overline{\gamma} (\overline{\gamma}+1) m_V^2}{\cosh^2(m_Vy)} \right] \tilde{\psi}^{(n)} 
  = \left(\overline{m_n}^2 - \overline{\gamma}^2 m_V^2 \right)  \tilde{\psi}^{(n)},  
\label{KK_EOM5}  
\eea
which is identical to the equation for $\tilde{\chi}^{(n)}$. 
Thus, the eigenvalues for $\tilde{\psi}^{(n)}$ are given by 
\bea 
  m_n^2 = (n+1) \left( 2 \gamma -(n+1) \right) m_V^2  \; \; \; \; (n=0, 1, 2, \cdots < \gamma-1).
\eea 
As for the first example, no zero-mode exists for the scalar mode, 
  and we can see the pairing of the KK-mode mass spectrum between the gauge fields 
    and the corresponding would-be NG modes.   
The eigenfunctions for even numbers of 
  $n= 2 n^\prime \; (n^\prime =0,1,2, \ldots )$ and odd numbers of 
  $n= 2 n^{\prime \prime}+1 \; (n^{\prime \prime} =0,1,2,\ldots )$ 
are given by (up to normalization factor) 
\begin{eqnarray} 
 \tilde{\psi}^{(n^\prime)}(y) = \left[ \cosh(m_V y) \right]^{- \gamma+1} \, 
   F\left[-n^\prime , -\gamma + n^\prime+1; 1/2 ; 1- \cosh^2(m_V y) \right], 
\end{eqnarray}
and 
\begin{eqnarray}
  \tilde{\psi}^{(n^{\prime \prime})}(y) =  \sinh(m_V y) 
      \left[\cosh(m_V y) \right]^{- \gamma+1}  
  F\left[-n^{\prime \prime}, - \gamma + n^{\prime \prime}+2 ; 3/2 ; 1- \cosh^2(m_V y) \right],   
\label{F2}
\end{eqnarray}
respectively.

Unlike the first example, we have a finite number of the localized KK-modes in the second example. 
For concreteness, let us fix $\gamma=2$. 
In this case, we have only one KK-mode gauge boson with the mass eigenvalue $m_1^2=3 m_V^2$.   
In the 4D effective theory with the canonically normalized kinetic terms like Eq.~(\ref{L_eff}), 
  the explicit form of the KK-mode expansions is given by
\bea
  A_{\mu}(x,y) &=& g \, A_{\mu}^{(0)}(x) + \sqrt{2} \, g \, \sinh(m_V y)  \, A_{\mu}^{(1)}(x), \nonumber \\
  A_{y}(x,y) &=&  \sqrt{\frac{2}{3}} \, g \, \cosh(m_V y)  \, \, \eta ^{(1)}(x),
\label{A_sol}  
\eea  
where the gauge coupling in the 4D effective theory is defined as $g=\sqrt{\frac{3 m}{4 M}}$.

It is straightforward to extend the U(1) gauge theory to the SM case. 
For the SM gauge group of SU(3)$_c\times$SU(2)$_L\times$U(1)$_Y$, 
  we introduce three $y$-dependent gauge couplings in the original 5D Lagrangian. 
Let us call them as $s_i(y)$ for $i=1, 2, 3$ corresponding to the three gauge groups. 
For simplicity, we set $s_1 \propto s_2$ in this paper, so that the KK-mode spectrum 
  for the SU(2)$_L\times$U(1)$_Y$ gauge bosons are the same. 
As we will see in the next section, this choice simplifies our calculation 
  of the KK-mode mass spectrum after the electroweak symmetry breaking.

%%%%%%%%%%%%%%%%%%%%%%%%    
\section{The Higgs sector} 
%%%%%%%%%%%%%%%%%%%%%%%%
\label{sec:3}

Next we consider the 5D Higgs sector of the Abelian Higgs model, 
  corresponding to the previous section on the localized U(1) gauge field.  
It is straightforward to extend our discussion to the SM Higgs doublet case. 
In a non-compact 5th dimension, we need to consider a localization mechanism 
  for not only Higgs field but also its vacuum expectation value (VEV). 
For this purpose, we apply the same procedure taken for the gauge field 
  in the previous section. 
We thus consider the Lagrangian for the Higgs sector of the form, 
\bea
\mathcal{L}_5^H=s_H(y) \left[ ({\cal D}^{M}H)^{\dagger}({\cal D}_{M}H)
  -\frac{1}{2}\lambda_H \left(H^{\dagger}H-\frac{v^2}{2} \right)^2 \right], 
\label{L_H}  
\eea
where $H$ is the Higgs field, $v$ is its VEV, $\lambda_H$ is a Higgs quartic coupling, 
  and the covariant derivative is given by ${\cal D}_M= \partial_M - i Q_H A_M$ 
  with a U(1) charge $Q_H$ for the Higgs field.   
Here, we have introduced a $y-$dependent kinetic term $s_H(y)$. 
In our convention, the Higgs field and $s_H$ have mass dimension one.

Expanding about the vacuum $H=(v + h + i \phi)/\sqrt{2}$ and neglecting the interaction terms, 
  we obtain (up to total derivative terms)
\bea
\mathcal{L}_5^H & \supset & \frac{1}{2} s_H \left[ (\partial^{M}h)(\partial_{M}h)-m_{h}^{2}h^2\right]
 +\frac{1}{2} s_H (\partial^{M}\phi)(\partial_{M}\phi) \nonumber \\
&=&-\frac{1}{2} s_H h(\Box_{4}+m_{h}^{2})h+\frac{1}{2}h\,\partial_y(s_H \partial_{y}h) 
  -\frac{1}{2} s_H \phi\Box_{4}\phi+\frac{1}{2}\phi \,\partial_{y}(s_H \partial_{y}\phi) , 
\eea
where $m_h^2= \lambda_H v^2$ is the physical Higgs boson mass. 
Applying the KK-mode decomposition to these fields, 
\bea
h(x,y)=\sum_{n=0}^{\infty}h^{(n)}(x) \, \chi^{(n)}_{h}(y), 
  \quad 
\phi(x,y)=\sum_{n=0}^{\infty}\phi^{(n)}(x) \, \chi^{(n)}_{\phi}(y), 
\eea
   we see that the KK-mode equations for $\chi^{(n)}_{h}$ and $\chi^{(n)}_{\phi}$ 
   are identical to that of the gauge boson in Eq.~(\ref{KK_EOM}) with the replacement by $s(y) \to s_H(y)$.    
Since the zero-mode $\phi^{(0)}$ is the would-be NG mode eaten 
   by $A_\mu^{(0)}$, the theoretical consistency requires the configurations of $\phi^{(0)}$ and $A_\mu^{(0)}$ to be identical.  
With the solutions of the KK-mode equations, the free Lagrangian for the scalar fields 
  in the 4D effective theory is described as  
\bea
\mathcal{L}_4^H \supset 
 - \frac{1}{2} \sum_{n=0}^{\infty}  \left[ \int_{-\infty}^\infty dy \, s_H \left(\chi_h^{(n)} \right)^2 \right] 
\left[
  h^{(n)}\left( \Box_{4} + (m_h^2 +m_n^2)\right) h^{(n)}  
+ \phi^{(n)} (\Box_{4} + m_n^2)  \phi^{(n)}     
\right],  
\eea 
where we have used $\chi^{(n)}_{h}(y) = \chi^{(n)}_{\phi}(y)$ since their equations are identical.  
After canonically normalizing kinetic terms, the KK-mode decomposition with respect to physical Higgs bosons 
  is formally given by   $h(x,y)=  h^{(0)}(x) +  \chi_h^{(n)}(y) \, h^{(n)}(x)$, 
where we have scaled the KK-mode functions so as to satisfy
\bea
 \int_{-\infty}^\infty dy \, s_H \left( \chi_h^{(n)} \right)^2 =1\;\;\;\;\; (n = 0,1,2...).
\eea
Recall that $ \chi_h^{(0)}$ is a constant. 
The U(1) gauge symmetry is broken by $\langle H \rangle =v/\sqrt{2}$, 
  from which the masses for the U(1) gauge bosons in the 4D effective theory are generated. 
Their mass terms are given by 
\bea 
\mathcal{L}_4^H \supset 
 \frac{1}{2} Q_H^2 v^2   \sum _{n,m = 0}^\infty  \left[ \int_{-\infty}^\infty dy \, s_H(y) \chi^{(n)}(y) \chi^{(m)}(y) \right]
  \eta^{\mu \nu} A_{\mu}^{(n)}(x)   A_{\nu}^{(m)}(x). 
\eea    
This formula indicates that the KK-mode gauge bosons, $A_{\mu}^{(n)}$ and $A_{\mu}^{(m)}$ ($n \neq m$), 
   have a mixing mass in general, and the analysis of the gauge boson mass spectrum is complicated.
For simplicity, let us take $s_H(y)  \propto s(y)$ in this paper, so that no mixing mass is generated 
   because of the orthogonal condition, 
\bea
     \int_{-\infty}^\infty dy \, s_H(y) \chi^{(n)}(y) \chi^{(m)}(y) \propto \int_{-\infty}^\infty dy \, s(y) \chi^{(n)}(y) \chi^{(m)}(y)=0,  
\eea    
   for $n \neq m$.   
After normalizing the kinetic terms for the zero-mode and KK-mode gauge bosons,  
    we find the mass spectrum as 
\bea
  m_A^{(0)} = Q_H  \, g \, v,  \quad m_A^{(n)} = \sqrt{m_n^2 + \left( m_A^{(0)} \right)^2}, 
\eea    
 for the zero-mode and the KK-modes ($n=1, 2, \cdots$), respectively. 
Here, we have obtained the zero-mode gauge boson mass 
  of the same form as the one in the Abelian Higgs model in 4D.

We now extend the model to the SM case. 
As we mentioned in the previous section, we set $s_1 \propto s_2$, 
  so that the KK-mode spectrum of the SU(2)$_L \times$U(1)$_Y$ gauge bosons are the same. 
The Higgs field in the Abelian Higgs model is extended to the SM Higgs doublet field. 
After the electroweak symmetry breaking  we have the gauge boson (photon ($\gamma$), $W$ boson and $Z$ boson) mass spectrum:   
  for the zero-modes, 
\bea
   m_\gamma =0, \; \; m_W = \frac{1}{2} g_2 v, \; \; m_Z = \frac{1}{2} g_Z v,    
\eea  
where $g_Z=\sqrt{g_2^2 + g_Y^2}$ with $g_2$ and $g_Y$ being the SU(2)$_L$ and U(1)$_Y$ gauge couplings, respectively, 
   and $v_h=246$ GeV is the Higgs doublet VEV, 
   while their KK-mode mass spectrum is given by   
\bea
   m_\gamma^{(n)} = m_n, \; \; m_W^{(n)} = \sqrt{m_n^2 + m_W^2}, \; \; m_Z^{(n)} = \sqrt{m_n^2 + m_Z^2}. 
\eea

%%%%%%%%%%%%%%%%%%%%%
\section{Domain-wall fermions} 
%%%%%%%%%%%%%%%%%%%%% 
\label{sec:4}
In this section, we consider localized chiral fermions, whose zero-modes 
   are identified with the SM chiral fermions. 
Again, we consider the U(1) gauge theory to simplify our discussion, 
  which can be easily extended to the 5D SM case.  
We follow a mechanism in Ref.~\cite{rubakov}
  to generate the domain-wall fermion in 5D space-time and first introduce 
  a real scalar field ($\varphi(x, y)$) in the 5D bulk: 
\begin{eqnarray} 
 {\cal L}_{(5)} = 
  \frac{1}{2} \left(\partial_{M} \varphi \right) \left( \partial^{M} \varphi \right) - V(\varphi)   \; ,  
\end{eqnarray}
where the scalar potential is give by 
\bea 
 V(\varphi) = \frac{m_\varphi^4}{2 \lambda} 
	   -m_\varphi^2 \varphi^2  
           + \frac{\lambda}{2} \varphi^4 .  
\eea

It is well known that there is a non-trivial background configuration $\varphi_0 (y)$ 
 as a solution of the equation of motion, 
 namely, the so-called kink solution,
\bea
\varphi_{\rm kink} (y) = \frac{m_\varphi}{\sqrt{\lambda}} 
\tanh [m_\varphi y] .  
\label{kink} 
\eea
Here, we have chosen the kink center at $y=0$, for simplicity. 
Expanding the scalar around the kink background, 
  $\varphi (x, y) = \varphi_{\rm kink}(y) + U_\varphi(y) \tilde{\varphi}(x)$, 
  we can solve the linearized equation of motion. 
It is easy to notice that this system is the same as the second solvable example in Sec.~\ref{sec:2} 
  with $\gamma=2$, so that we have the solution \cite{Dashen}: 
\bea
   \varphi (x, y) = \varphi_{\rm kink}(y) +  \frac{\sqrt{3 m_\varphi} }{2} \left[ \frac{1}{\cosh^2(m_\varphi y)} \right] \varphi^{(0)}(x) 
    + \sqrt{ \frac{3m_\varphi}{2} } \left[ \frac{\sinh (m_\varphi y)}{\cosh^2 (m_\varphi y)} \right] \varphi^{(1)}(x), 
\label{s-mode}    
\eea  
where $\varphi^{(0)}(x)$ is a massless NG mode corresponding to the spontaneous breaking 
  of the translational invariance in the 5th dimension, 
  and $\varphi^{(1)}(x)$ is a massive mode with a mass $m_\varphi^{(1)} = \sqrt{3} m_\varphi$.  
Here, the kinetic terms for the eigenstates are canonically normalized.

Following Ref.~\cite{rubakov}, we now introduce the Lagrangian for a bulk fermion coupling with $\varphi$, 
\bea
\mathcal{L}&=&i \overline{\psi}\left[\gamma^{\mu}D_{\mu}+i\gamma^{5}D_{y}\right]\psi+Y \varphi \overline{\psi}\psi \nonumber \\ 
&=& 
i \overline{\psi_L} \gamma^{\mu}D_{\mu} \psi_{L}+i \overline{\psi_R} \gamma^{\mu}D_{\mu} \psi_{R}  \nonumber \\
&-& \overline{\psi_L} D_{y}\psi_{R}+\bar{\psi}_{R}D_{y} \psi_L + Y \varphi \left( \overline{\psi_L}\psi_{R}+ \overline{\psi_R}\psi_{L}\right), 
\label{DM_fermion}
\eea
where we decompose the Dirac fermion $\psi$ into its chiral components, $\psi= \psi_L + \psi_R$, 
  the covariant derivative is given by $D_M=\partial_M - i Q_f A_M$ with a U(1) charge $Q_f$ for $\psi$, 
  and $Y$ is a positive constant.   
Neglecting the gauge interaction and replacing $\varphi$ by the kink solution, 
  the equations of motion are given by 
\bea
&& i\gamma^{\mu} \partial_{\mu}\psi_{L}- \partial_{y}\psi_{R} + Y \varphi_0 \psi_{R}=0,  \nonumber \\
&& i\gamma^{\mu}  \partial_{\mu}\psi_{R}+ \partial_{y}\psi_{L} + Y \varphi_0 \psi_{L}=0.
\eea
Using the KK-mode decompositions, 
\bea
 \psi_L (x,y) = \sum_{n=0}^\infty \psi_L^{(n)}(x) \, \chi_L^{(n)}(y), \; \; \;
 \psi_R (x,y) =\sum_{n=0}^\infty \psi_R^{(n)}(x) \, \chi_R^{(n)}(y), 
\eea
  we have the KK-mode equations as 
\bea
& \left[ -\partial_y^2 - (Y \varphi_0)^\prime  +  (Y \varphi_0)^2 \right]  \chi_L^{(n)} = m_n^2   \chi_L^{(n)}, \nonumber\\
& \left[ -\partial_y^2 + (Y \varphi_0)^\prime  +  (Y \varphi_0)^2  \right] \chi_R^{(n)} = m_n^2  \chi_R^{(n)}. 
\label{KKF_EOM}
\eea
We can easily show that these equations are equivalent to the two equations in Eq.~(\ref{KK_EOM4}) 
   by the identifications, 
   $m_\varphi \to m_V$, $Y/\sqrt{\lambda} \to \gamma$, and $\chi_L^{(n)}, \, \chi_R^{(n)} \to \tilde{\chi}^{(n)}, \, \tilde{\psi}^{(n)}$.
Hence, the mass eigenvalues and eigenfunctions are given by Eqs.~(\ref{mn1})-(\ref{F2}).   
Note that we have only one (left) chiral fermion in the 4D effective theory, 
   which is identified with an SM fermion in the extension of the SM case.

For our phenomenology discussion in the next section, let us set $Y/\sqrt{\lambda} =2$. 
In this case, we have only one KK-mode Dirac fermion in the 4D effective theory 
  with mass $\sqrt{3} m_\varphi$. 
The KK-mode expansion is explicitly described as 
\bea
 \psi_L (x,y) &=& 
  \frac{\sqrt{3 m_\varphi} }{2} \left[ \frac{1}{\cosh^2(m_\varphi (y-y_0))} \right] \psi_L^{(0)}(x)
    + \sqrt{ \frac{3m_\varphi}{2} } \left[ \frac{\sinh (m_\varphi (y-y_0))}{\cosh^2 (m_\varphi (y-y_0))} \right] \psi_L^{(1)}(x),   \nonumber \\
 \psi_R (x,y) &=&  \sqrt{ \frac{m_\varphi}{2}} \left[ \frac{1}{\cosh(m_\varphi (y-y_0))} \right] \psi_R^{(1)}(x), 
\label{F_example} 
\eea
for which the kinetic terms are canonically normalized. 
Here, we have generalized the system and set the kink center at $y_0$. 
The KK-mode functions for $\psi_L$ are the same as those for the scalar $\varphi$  
  shown in Eq.~(\ref{s-mode}).

%%%%%%%%%%%%%%%%%%%%%%%%%%%%%%%%
\begin{figure}[t]
\begin{center}
\includegraphics[scale=0.9]{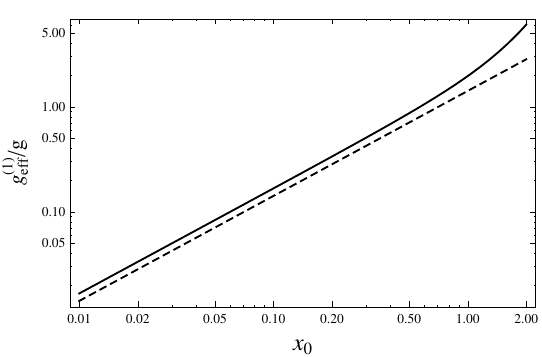} \;
\includegraphics[scale=0.9]{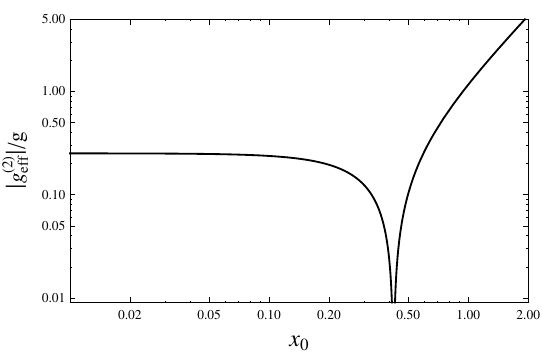} 
\end{center}
\caption{
Left: The effective coupling ($g_{\rm eff}^{(1)}$) between the 1st KK-mode gauge boson and the chiral fermion 
  as a function of $x_0$ for the first (solid) and second (dashed) examples in Sec.~\ref{sec:2}.
The gauge coupling of the zero-mode gauge boson is denoted as $g$.   
Right: The effective coupling ($g_{\rm eff}^{(2)}$) of the 2nd KK-mode gauge boson 
  with the chiral fermion as a function of $x_0$ for the second example in Sec.~\ref{sec:2}.
For $x_0 > 0.421$, we find $g_{\rm eff}^{(2)} < 0$.   
}
\label{fig:1}
\end{figure}
%%%%%%%%%%%%%%%%%%%%%%%%%%%%%%%%%%

Let us now describe the Lagrangian for the chiral fermion in the 4D effective theory as 
\bea
 {\mathcal L}_4 \supset  
  \overline{\psi_L^{(0)}} i \gamma^\mu \left( \partial_\mu - i Q_f g A_\mu^{(0)} \right) \psi_L^{(0)} 
+ \sum_{n=1}^\infty  Q_f \, g_{\rm eff}^{(n)} \, A_\mu^{(n)}  \left[\overline{\psi_L^{(0)}} \gamma^\mu \psi_L^{(0)} \right], 
\eea
where the 4D effective gauge coupling between the chiral fermion and the $n$-th KK-mode gauge boson is given by  
\bea
   g_{\rm eff}^{(n)} =  \int_{-\infty}^{\infty} dy  \left( \chi_{L}^{(0)} \right)^2  \chi^{(n)} .
\eea
For simplicity, we consider the solvable examples in Sec.~\ref{sec:2} and take $m_V=m_\varphi$. 
For the first and second examples, the effective gauge couplings of the 1st KK-mode gauge boson are, respectively, given by
\bea 
  {\rm Example~1}:\;  \frac{g_{\rm eff}^{(1)}}{g} &=& \frac{3 \sqrt{2}}{4} \, m_\varphi^2
     \int_{-\infty}^{\infty} dy  \,  \frac{y}{\cosh^4(m_\varphi (y-y_0))} \nonumber \\
   &=&   \frac{3 \sqrt{2}}{4} \, 
     \int_{-\infty}^{\infty} dx  \,  \frac{x}{\cosh^4(x-x_0)},  \nonumber \\
 {\rm Example~2}: \; \frac{g_{\rm eff}^{(1)}}{g} &=& \frac{3 \sqrt{2}}{4} \, m_\varphi 
     \int_{-\infty}^{\infty} dy  \,  \frac{\sinh(m_\varphi y)}{\cosh^4(m_\varphi (y-y_0))}\nonumber \\
   &=&   \frac{3 \sqrt{2}}{4} \, 
     \int_{-\infty}^{\infty} dx  \,  \frac{\sinh(x)}{\cosh^4(x-x_0)},  
\label{g_eff_formula}     
\eea
where $x_0=m_\varphi y_0$. 
Since the eigenfunction of the 1st KK-mode gauge boson is an odd-function of $y$, 
  the effective coupling vanishes $g_{\rm eff}^{(1)}/{g} \to 0$ for $x_0 \to 0$. 
In the first example, there is an infinite tower of KK-modes, and we also calculate 
  the effective gauge coupling of the 2nd KK-mode gauge boson, 
\bea
\frac{g_{\rm eff}^{(2)}}{g} = \frac{3}{4 \sqrt{2}} \, m_\varphi 
     \int_{-\infty}^{\infty} dy  \,  \frac{1- 2 (m_\varphi y)^2}{\cosh^4(m_\varphi (y-y_0))}
   =   \frac{3}{4 \sqrt{2}} \, 
     \int_{-\infty}^{\infty} dx  \,  \frac{1- 2 x^2}{\cosh^4(x-x_0)}, 
\eea  
which approaches $g_{\rm eff}^{(2)}/g \to 0.251$ for $x_0 \to 0$.

In Figure \ref{fig:1}, we show the effective gauge couplings between the 1st KK-mode gauge boson 
   and the chiral fermion for the first (solid) and second (dashed) examples (left panel), 
   and the effective gauge coupling of the 2nd KK-mode gauge boson 
   with the chiral fermion (right panel). 
In the left panel, the gauge couplings vanish for $x_0 \to 0$, 
   while  the gauge coupling of the 2nd KK-mode approaches a constant value, $g_{\rm eff}^{(2)}/g \to 0.251$. 
We find $g_{\rm eff}^{(2)} < 0$ for $x_0 > 0.421$.  
When applied to the SM, the gauge coupling $g$ corresponds to one of the SM gauge couplings 
  and the chiral fermion is identified with an SM fermion.  
We will discuss implications of this coupling behavior to LHC phenomenology in the next section.

Finally, let us extend our system to the SM case, and we introduce 
  the Yukawa coupling of the SM fermions in 5D as 
\bea
  {\mathcal L}_Y = - Y_f \overline{D} H S +{\rm H.c.} 
  =- Y_f \overline{D_L} H S_R  - Y_f \overline{D_R} H S_L + {\rm H.c.}, 
\label{Yukawa}  
\eea
where $D$ and $S$ are 5D fermions of the SM SU(2) doublet and singlet, respectively, 
  we have decomposed them into their chiral components $D=D_L+D_R$ and $S=S_L+S_R$, 
  and  $H$ is the 5D Higgs doublet. 
With the kink background, zero-modes of $D_L$ and $S_R$ are identified 
  with left-handed SM doublet and right-handed singlet fermions. 
For simplicity, suppose the KK-mode expansions for $D$ and $S$ (we exchange the chiraliteis for $S$) 
  are given by Eq.~(\ref{F_example}). 
For the Higgs doublet field in 5D, let us take, for simplicity, 
\bea 
   s_H(y)=\frac{3}{4} \,  \frac{m_\varphi}{\cosh^4 (m_\varphi y)}
\eea  
  as in Sec.~\ref{sec:2}, so that the KK-mode decomposition 
  of the physical Higgs boson after the electroweak symmetry breaking are given by  
  (see Eq.~(\ref{A_sol}))
\bea
  h(x,y)=  h^{(0)}(x)  + \sqrt{2} \sinh(m_\varphi y) \, h^{(1)}(x).   
\label{H_KK}
\eea  
When we identify Eq.~(\ref{Yukawa}) with top Yukawa coupling in 5D,  
  we obtain a 4D effective Yukawa coupling for the top quark as 
\bea
  {\mathcal L}_Y^4  \supset - m_t \left(1+\frac{h^{(0)}}{v} \right)\overline{t_L} t_R
  - \left(\frac{m_t}{v} \right) C_{\rm eff} \,  h^{(1)} \, \overline{t_L} t_R, 
\label{Yukawa_KK_Higgs}  
\eea
where $m_t=Y_f v/\sqrt{2}$, and $C_{\rm eff}$ is given by Eq.~(\ref{g_eff_formula}).
The top quark mass formula is the same as the SM one in 4D, while the model involves  
  a KK-mode Higgs boson with the Yukawa coupling, $m_t C_{\rm eff}/v$ 
  and its mass $\sqrt{3 m_\varphi^2+ m_h^2}$ with the SM Higgs boson mass $m_h$.

%%%%%%%%%%%%%%%%%%%%%
\section{KK-mode Phenomenology}
%%%%%%%%%%%%%%%%%%%%%
\label{sec:5}

Prediction of the KK-modes in the 4D effective theory is a common property 
  of extra-dimensional models and  we can investigate the phenomenology involving the KK-modes. 
However, in the Domain-Wall SM, the KK-mode spectra and the coupling manner of each KK-mode 
  with the SM particles depend on the localization mechanism. 
This property is in sharp contrast to, for example, the Universal Extra-Dimension model \cite{UED}, 
  where the KK-mode eigenfunctions are uniquely determined by boundary 
  conditions associated with the compactification of the 5th dimension.  
The Domain-Wall SM offers more variety of the KK-mode phenomenologies 
  than usual compactified extra-dimensional models, 
  thanks to rich geometrical structures of the localized SM particles and their KK-modes. 
In this section, we address a few directions for interesting KK-mode phenomenologies.

%%%%%%%%%%%%%%%%%%%%%%%%%%%%%%
\subsection{Phenomenology of KK-mode gauge bosons} 
%%%%%%%%%%%%%%%%%%%%%%%%%%%%%%

The ATLAS and the CMS collaborations have been searching for a new gauge boson resonance 
  with a variety of final states at the LHC Run-2.  
The so-called sequential SM $Z^\prime$ and $W^\prime$ bosons, 
  which have the same properties as the SM $Z$ and $W$ bosons except for their masses, 
  have been examined as a benchmark model. 
According to the LHC Run-2 final report by the ATLAS collaboration on their search results 
  with $\sqrt{s}=13$ TeV and an integrated luminosity of 139 fb$^{-1}$, 
  the lower mass bound on the sequential SM $Z^\prime$ boson is $m_{Z^\prime} \geq 5.1$ TeV, 
  which is obtained by the search with dilepton final states \cite{ATLAS:2019} 
  (a similar bound, $m_{Z^\prime} \geq 5.15$ TeV, is obtained by the CMS collaboration \cite{CMS:2019}). 
A more severe constraint, $m_{W^\prime} \geq 6.0$ TeV, is obtained for the sequential SM $W^\prime$ boson mass 
  from the search with its decay mode $W^\prime \to l \nu$  \cite{ATLAS2:2019} 
  (a similar bound,  $m_{W^\prime} \geq 5.7$ TeV, is obtained by the CMS collaboration \cite{CMS2:2022}).  
In this subsection, we interpret these results as constraints on the KK-mode gauge bosons in the Domain-Wall SM.

In Sec.~\ref{sec:2}, we have chosen $s_1 \propto s_2$, 
  so that the mass spectrum of the KK-mode $Z$ and $W$ bosons are the same, 
  neglecting the mass terms generated by the electroweak symmetry breaking.  
Thus, we consider the most severe constraint from the $W^\prime$ boson search. 
Since the total decay width of the sequential $W^\prime$ boson is about 3\% of its mass for $m_{W^\prime} \gtrsim 1$ TeV, 
   we employ the narrow-width approximation in evaluating the parton-level cross section of the process, 
\bea 
   \hat{\sigma}(q \overline{q^\prime} \to W^\prime) \propto \Gamma_{W^\prime}(W^\prime \to q \overline{q^\prime})  
      \, \delta(M_{\rm inv}^2 -m_{W^\prime}^2)  \propto  g^2,  
\eea   
where $M_{\rm inv}^2$ is the invariant mass of the initial partons, 
   $\Gamma_{W^\prime}(W^\prime \to q \overline{q^\prime})$ is the partial decay width into $q \overline{q^\prime}$, 
   and $g$ is the SM SU(2) gauge coupling. 
The difference of the sequential $W^\prime$ boson and the KK-mode $W$ boson in the Domain-Wall SM 
   is only the coupling constant. 
Because of the non-trivial eigenfunctions of the SM fermions and the KK-mode $W$ boson, 
   the effective gauge coupling constant is not the same as the SM gauge coupling constant, 
   as shown in Figure \ref{fig:1} for our example.     
Hence, we have a relation between the production cross sections of the sequential $W^\prime$ 
   and the 1st KK-mode $W$ bosons in the narrow-width approximation:  
\bea
   \sigma (pp \to W^{(1)} \to l \nu) = \left( \frac{g_{\rm eff}^{(1)}}{g} \right)^2 \sigma (pp \to W^\prime \to l \nu),  
\eea
  by which we can interpret the current ATLAS constraints as those on the 1st KK-mode $W$ boson ($W^{(1)}$).

%% %%%%%%%%%%%%%%%%%%%%%%%%%%%%%%%%
\begin{figure}[t]
\begin{center}
\includegraphics[scale=0.5]{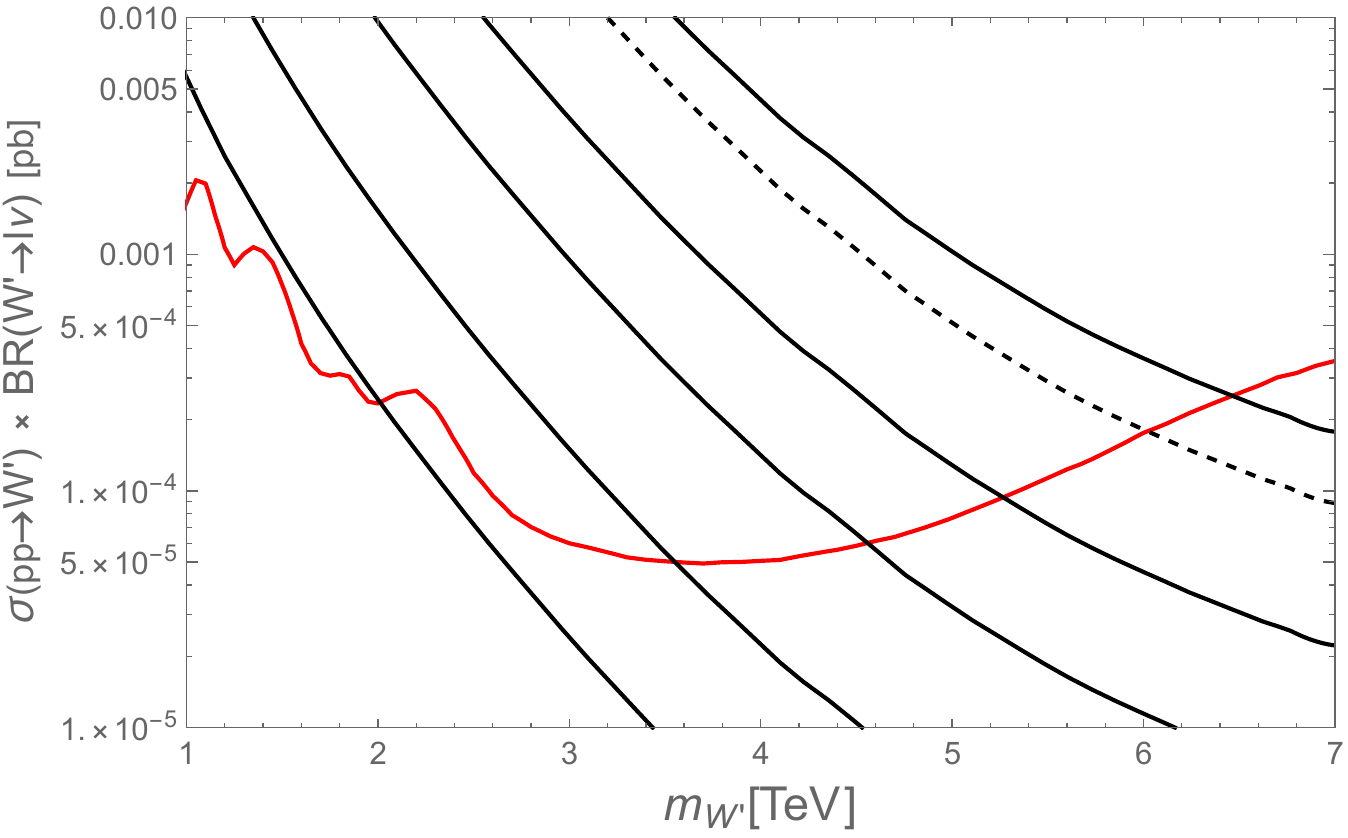}       
\end{center}
\caption{
The cross section $\sigma (pp \to W^{(1)} \to l \nu)$ as a function of $m_{W^\prime}=m_{W^\prime}^{(1)}$
  for $g_{\rm eff}^{(1)}/g=0.04$, $0.1$, $0.251$, $0.5$ and $\sqrt{2}$ (solid lines) from left to right, 
  along with the upper bound on the cross section from the ATLAS results (red horizontal curve)
  and the theoretical prediction of $\sigma (pp \to W^\prime \to l \nu)$ for the sequential SM $W^\prime$ boson (dashed line). 
For these $g_{\rm eff}^{(1)}/g$ values, we find the lower bounds on the 1st KK-mode $W$ boson as 
   $m_{W^\prime}^{(1)}[{\rm TeV}]  \geq 2.0$, $3.5$, $4.6$ $5.3$ and $6.5$, respectively.  
The result for $g_{\rm eff}^{(1)}/g=0.251$ can be identified as the result for the 2nd KK-mode $W$ boson 
  in the limit of $x_0 \to 0$.  
}
\label{fig:2}
\end{figure}
%%%%%%%%%%%%%%%%%%%%%%%%%%%%%%%%%%

In Figure \ref{fig:2}, we show the cross section $\sigma (pp \to W^{(1)} \to l \nu)$ as a function of $m_{W^\prime}=m_{W^\prime}^{(1)}$
  for various values of $g_{\rm eff}^{(1)}/g$, 
  along with the upper bound on the cross section from the ATLAS results \cite{ATLAS2:2019} (horizontal curve (in red))
  and the theoretical prediction of $\sigma (pp \to W^\prime \to l \nu)$ for the sequential SM $W^\prime$ boson (dashed line). 
The solid diagonal lines from left to right depict the theoretical predictions 
  of the cross section $\sigma (pp \to W^{(1)} \to l \nu)$ 
  for $g_{\rm eff}^{(1)}/g=0.04$, $0.1$, $0.251$, $0.5$ and $\sqrt{2}$, respectively, 
  corresponding to 
  $x_0=0.024$, $0.060$, $0.15$, $0.30$ and $0.77$ ($x_0=0.028$, $0.071$, $0.18$, $0.35$ and $1.0$) 
  from the solid (dashed) line in the left panel of Figure \ref{fig:1}. 
For these $g_{\rm eff}^{(1)}/g$ values, we read off the lower bounds on the 1st KK-mode mass as   
  $m_{W^\prime}^{(1)}[{\rm TeV}]  \geq 2.0$, $3.5$, $4.6$ $5.3$ and $6.5$, respectively.  
The result for $g_{\rm eff}^{(1)}/g=0.251$ is identified with the result for the 2nd KK-mode gauge boson 
  for $x_0 \to 0$ from the right panel of Figure \ref{fig:1}.

The structure of the KK-modes depend on a localization mechanism, in particular, 
  the shape of $y$-dependent gauge couplings. 
We have introduced two solvable examples in Sec.~\ref{sec:2}: the first one predicts an infinite tower 
  of KK-modes, while the KK-mode expansion is terminated in the second example. 
The second example is in sharp contrast to extra-dimensional models with compactified extra-dimensions, 
   which predict the infinite tower of KK-mode spectrum. 
Once the 1st KK-mode state is discovered, 
   the search for higher KK-mode states at high energy colliders can test 
   if an extra-dimension is compactified or not.

%%%%%%%%%%%%%%%%%%%%%%%%%%%%%%
\subsection{Higgs boson phenomenology} 
%%%%%%%%%%%%%%%%%%%%%%%%%%%%%%

Since the KK-modes of the SM particles have couplings with the SM Higgs boson, 
  the presence of the KK-modes affects on Higgs boson phenomenology.  
In particular, the Higgs boson properties measured at the LHC \cite{PDG}
  are altered from the SM predictions.  
In this section, we consider implication of the KK-modes of top quark and  $W$ boson 
  to the Higgs boson phenomenology.  
See, for example, Ref.~\cite{Effective_Higgs_Cooupling}, for a pioneering work of this direction.

The SM Higgs boson has effective couplings with digluon and diphoton of the form, 
\bea 
 {\cal L}_{\rm Higgs-gauge} = 
  C_{gg} \, h \, G^A_{\mu \nu} G^{A \mu \nu}  + C_{\gamma \gamma} \, h \, F_{\mu \nu} F^{\mu \nu},   
 \label{Higgs-gauge}
\eea
where $G^A_{\mu \nu}$ $A=1,2, \cdots, 8$  and $F_{\mu \nu}$ are the field strengths of gluon and photon. 
In the SM, these effective operators are induced dominantly through 1-loop corrections 
   of top quark and $W$-boson (and associated would-be NG bosons and ghosts) \cite{HHG}. 
The effective coupling with digluon from top quark loop corrections is calculated to be 
\bea 
  C_{gg}^{\rm SM} = \frac{\alpha_{\rm s}}{16 \pi v} F_{1/2}(\tau_t), 
 \label{SM-glue}
\eea
where $\alpha_s$ is the QCD coupling, 
  and $\tau_t = 4 m_t^2/m_h^2$ with top quark mass $m_t$, and Higgs boson mass $m_h$.   
The effective coupling with diphoton is from 1-loop corrections with top quark and W-boson, 
  and  we have  
\bea 
 C_{\gamma \gamma}^{\rm SM} = 
  \frac{\alpha_{\rm em}} {8 \pi v} 
  \left( 
   \frac{4}{3} F_{1/2}(\tau_t) + F_{1}(\tau_W) 
  \right) ,     
\label{SM-gamma}
\eea 
where $\tau_W = 4 m_W^2/m_h^2$ with the $W$-boson mass $m_W$. 
The explicit formulas of the loop functions are given by  
\begin{eqnarray}
F_{1/2}(\tau) =
  2 \tau \left[ 1+ \left( 1 - \tau  \right) f(\tau) \right] , 
\; \; \; \; 
F_{1}(\tau) = 
 -\left[2 + 3 \tau + 3 \tau \left( 2 - \tau  \right) f(\tau) \right] , 
\end{eqnarray}
with $f(\tau)  =  \left[ \sin^{-1}\left( 1/\sqrt{\tau}\right)\right]^2$ ($ \tau > 1$). 
For $m_t=172.69.$ GeV, $m_W=80.377$, and $m_h=125.25$ GeV \cite{PDG}, 
  we find $F_{1/2}(\tau_t) \simeq 1.38$ and  $F_{1}(\tau_W) \simeq -8.33$.

In the presence of the KK-mode top quarks and $W$ bosons, the effective Higgs couplings  
   receive 1-loop corrections with the KK-modes. 
Again, we consider the KK-mode expansions in Eq.~(\ref{F_example}) for top quarks of the SM SU(2) doublet component 
   and the SU(2) singlet. 
After the electroweak symmetry breaking, it is easy to derive the mass matrix for the 1st KK-mode top quarks as 
\bea 
{\cal M}_t=
\left[
\begin{array}{cc}
   - m_t^{(1)} &  m_t  \\
   m_t  & m_t^{(1)} \\
\end{array}
\right], 
\eea
where $m_t^{(1)}=\sqrt{3} m_\varphi$ is the KK-mode mass. 
We then have degenerate mass eigenvalues, $m_{\rm KK}=\sqrt{  \left(m_t^{(1)} \right)^2 + m_t^2 }$. 
For $m_{\rm KK}^2 \gg m_h^2$, we can easily calculate the contribution of the KK-mode top quarks to $C_{gg}$
  by using the Higgs low-energy theorem \cite{HHG} as 
\bea
  C_{gg}^{\rm KK-top} \simeq \frac{\alpha_{\rm s}}{8 \pi v} b_3^t  \frac{\partial}{\partial \log v} \log (m_{\rm KK}) \times 2 
      \simeq \frac{\alpha_{\rm s}}{6 \pi v} \left( \frac{m_t}{m_{\rm KK}} \right)^2, 
\eea  
where $b^t_3=2/3$ is the top quark contribution to the beta function coefficient of QCD.
Similarly, the contribution to $C_{\gamma \gamma}$ from the KK-mode top quarks is given by 
\bea
 C_{\gamma \gamma}^{\rm KK-top} \simeq \frac{\alpha_{\rm em}}{6 \pi v} b_1^t  \frac{\partial}{\partial \log v} \log (m_{\rm KK}) \times 2 
      \simeq \frac{4 \alpha_{\rm em}}{9 \pi v} \left( \frac{m_t}{m_{\rm KK}} \right)^2, 
\eea  
where $b_t^t=4/3$ is a top quark contribution to the QED beta function coefficient. 
In calculating the contribution from the KK-mode $W$ boson, 
  we consider the expansion in Eq.~(\ref{A_sol}) with $m_V=m_\varphi$ to simplify our analysis.  
In this case, the 1st KK-mode $W$ boson has mass $\sqrt{  \left(m_W^{(1)} \right)^2 + m_W^2 } \simeq m_{\rm KK}$, 
  and we find the contribution to $C_{\gamma \gamma}$ from the KK-mode $W$ boson as 
\bea
 C_{\gamma \gamma}^{{\rm KK}-W} \simeq 
   \frac{\alpha_{\rm em}}{8 \pi v} b_1^W  \frac{\partial}{\partial \log v} 
   \log \left(\sqrt{  \left(m_W^{(1)} \right)^2 + m_W^2 } \right) 
      \simeq -\frac{7 \alpha_{\rm em}}{8 \pi v} \left( \frac{m_W}{m_{\rm KK}} \right)^2, 
\eea  
where $b_1^W=-7$ is the $W$-boson contribution to the QED beta function coefficient.

%%%%%%%%%%%%%%%%%%%%%%%%%%%%%%%%
\begin{figure}[t]
\begin{center}
\includegraphics[scale=0.4]{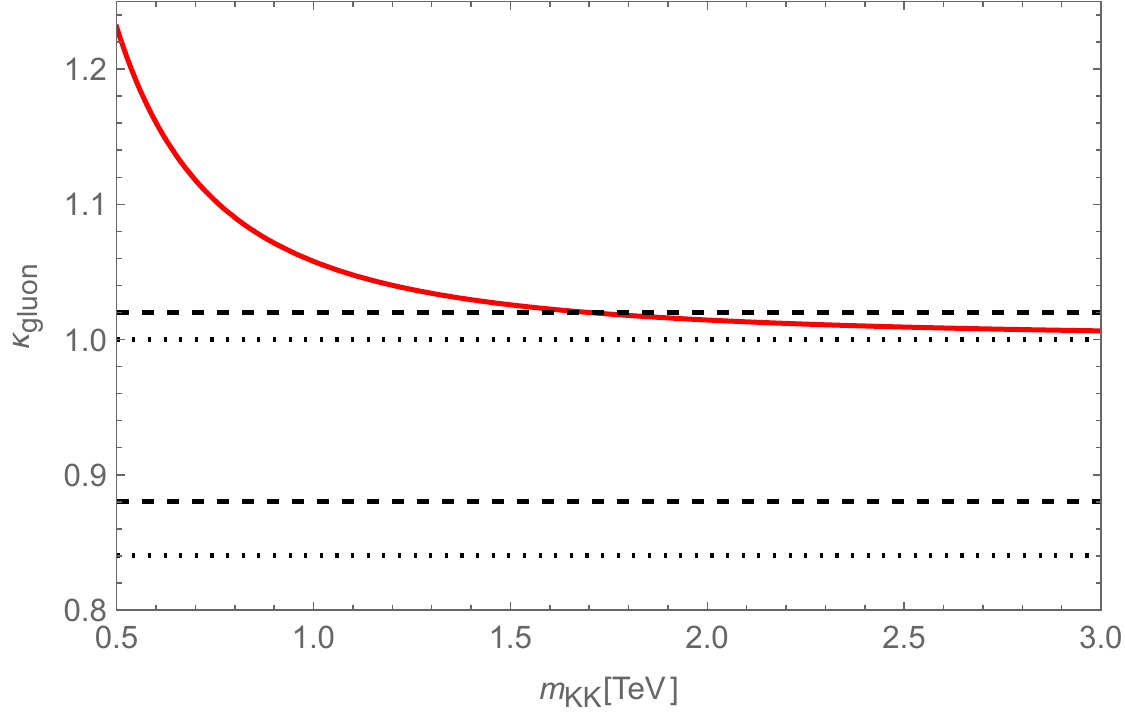} \;
\includegraphics[scale=0.4]{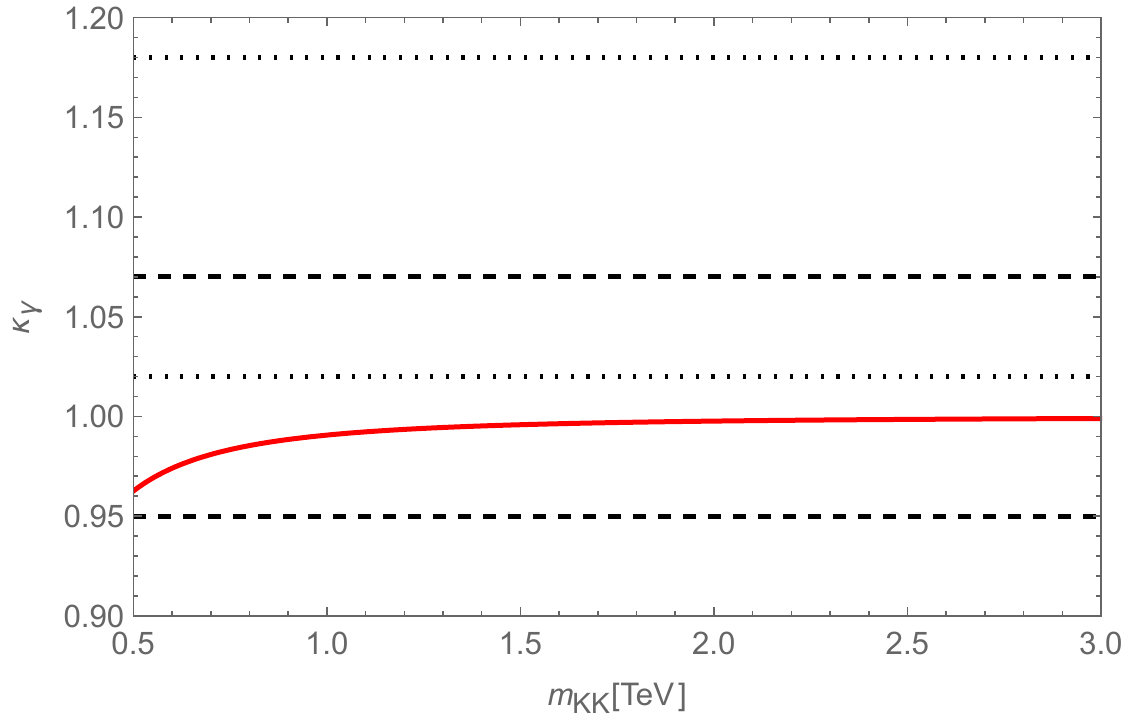} 
\end{center}
\caption{
Left: The ratio of the effective Higgs boson coupling with digluon to the SM one 
     as a function of $m_{\rm KK}$ (red solid line), along with the ATLAS and CMS results. 
Right: The ratio of the effective Higgs boson coupling with diphoton to the SM one 
     as a function of $m_{\rm KK}$ (red solid line), along with the ATLAS and CMS results. 
}
\label{fig:3}
\end{figure}
%%%%%%%%%%%%%%%%%%%%%%%%%%%%%%%%%%

At the LHC, the Higgs bosons are produced through various production processes
   and  several Higgs boson decay channels have been detected. 
The progress on precision measurements of the Higgs boson properties since the first discovery in 2012 
  has been reported by the ATLAS collaboration \cite{ATLAS:2022vkf} and the CMS collaboration \cite{CMS:2022dwd}. 
Their results show that the measured Higgs boson properties are  consistent with the SM predictions, 
  constraining the new physics contributions. 
  
Let us first evaluate the ratio of the effective Higgs boson coupling with digluon to the SM one as
\bea
   \kappa_{\rm gluon}  = 1 + \frac{C_{gg}^{\rm KK-top}}{C_{gg}^{\rm SM}}  . 
\eea 
This ratio as a function of $m_{\rm KK}$ is shown in Figure \ref{fig:3} (left panel). 
In the presence of the KK-mode top quarks, $\kappa_{\rm gluon}$ is deviating from $\kappa_g=1$ as $m_{\rm KK}$ is lowered. 
In the figure, we also show the ATLAS result \cite{ATLAS:2022vkf} of  $0.88 \leq  \kappa_{\rm gluon} \leq 1.02$
  at 1 $\sigma$ Confidence Level (region between two dashed lines) and the CMS result \cite{CMS:2022dwd} of 
  $0.84 \leq  \kappa_{\rm gluon} \leq $ at 1.00 $\sigma$ Confidence Level (region between two dashed lines).
We find the lower bound on $m_{\rm KK} \geq 1.7$ TeV to be consistent with the ATLAS result, 
  while our result is not consistent with the CMS result at the 1 $\sigma$ Confidence Level. 
The KK mode mass of $m_{\rm KK} \gtrsim 0.5 $ TeV is consistent with both of the ATLAS and CMS results 
  when we adopt 2-$\sigma$ confidence level.

We also evaluate the ratio of the effective Higgs boson coupling with diphoton to the SM one as
\bea
   \kappa_\gamma  = 1 + \frac{C_{gg}^{\rm KK-top}+C_{gg}^{\rm KK-W}}{C_{gg}^{\rm SM}}, 
\eea 
which is shown in Figure \ref{fig:3} (right panel), 
along with the ATLAS result \cite{ATLAS:2022vkf} of  $0.95 \leq  \kappa_g \leq 1.07$
  at 1 $\sigma$ Confidence Level (region between two dashed lines) and the CMS result \cite{CMS:2022dwd} of 
  $1.02 \leq  \kappa_g \leq 1.18$ at 1 $\sigma$ Confidence Level (region between two dashed lines).
We find our result for $m_{\rm KK} \gtrsim 0.5 $ TeV is consistent with the ATLAS result, 
  while our result is inconsistent with the CMS result at the 1 $\sigma$ Confidence Level.  
When we adopt the CMS result at 2-$\sigma$ confidence level, our result becomes consistent.

The SM Higgs boson is accompanied by a KK-mode Higgs boson which couple with a top quark pair, 
  as we have seen in Eqs.~(\ref{H_KK}) and (\ref{Yukawa_KK_Higgs}). 
Hence, the KK-mode Higgs boson can be produced at the LHC through the gluon fusion. 
The coupling of the KK-mode Higgs boson to top quarks 
  can be enhanced for a large $x_0$ (see the left panel of Figure \ref{fig:1}).  
It would be worth investigating the LHC phenomenology for the KK-mode Higgs boson.

%%%%%%%%%%%%%%%%%%%%%%%%%%%%%%
\subsection{Phenomenology of KK-mode fermions} 
%%%%%%%%%%%%%%%%%%%%%%%%%%%%%%

Let us consider interactions between the SM fermions and their KK-modes. 
Since the eigenfunctions of the SM gauge bosons, which are the zero-modes,  
  are independent of $y$, an interaction among one SM fermion, one KK-mode fermion and 
  one SM gauge boson vanishes by the orthogonal condition for the eigenfunctions. 
This is also true for a Yukawa interaction among one SM fermion, one KK-mode fermion 
  and one SM Higgs boson, if the SM SU(2) doublet and singlet fermions are decomposed 
  by the same KK-mode eigenfunctions, as we have considered in this paper 
  to simplify our discussions. 
However, there is a unique interaction between an SM fermion and its KK-mode 
  derived from the Yukawa interaction with the kink scalar in Eq.~(\ref{DM_fermion}),  
\bea
\mathcal{L} &\supset & Y \varphi \left( \overline{\psi_L}\psi_{R}+ \overline{\psi_R}\psi_{L}\right) \nonumber \\
  &\supset&  Y \, 
 \frac{3 m_\varphi}{2} \sqrt{\frac{m_\varphi}{2}} 
 \left[ \frac{1}{\cosh^5(m_\varphi (y-y_0))} \right]  \varphi^{(0)}(x) \overline{\psi_L^{(0)}(x)} \psi_R^{(1)}(x) + {\rm H.c.}, 
\eea
where we have used the KK-mode expansions in Eqs.~(\ref{s-mode}) and (\ref{F_example}) 
   with the shift of the kink center to $y=0 \to y_0$.  
Integrating out the 5th dimensional degrees of freedom, we obtain a 4D effective interaction, 
\bea
\mathcal{L}_{\rm eff}   \supset  y_{\rm eff} \,  \varphi^{(0)} \overline{\psi_L^{(0)}} \psi_R^{(1)}+ {\rm H.c.} , 
\label{Yukawa_NG}
\eea
where 
\bea
y_{\rm eff}  = \frac{3 m_\varphi}{2} \sqrt{\frac{m_\varphi}{2}} 
 \, \int_{-\infty}^\infty  \frac{dy }{\cosh^5(m_\varphi (y-y_0))} 
 = Y \,  \frac{9 \pi}{16}  \sqrt{\frac{m_\varphi}{2}} . 
 \eea
Therefore, the KK-mode fermion decays to the zero-mode fermion and
  the massless scalar, the NG boson associated with the breakdown 
  of 5th dimensional translational invariance.

In the extension to the SM case, we identify the fermions in Eq.~(\ref{Yukawa_NG}) with an SM fermion and it's ``KK-mode partner.''
This Yukawa interaction leads to an interesting phenomenology for the KK-mode fermions. 
At the LHC, a pair of KK-mode fermions can be produced through gauge interactions. 
For example, we may consider a pair of KK-mode quarks produced through gluon fusion. 
Once a KK-mode fermion is produced, it decays into an SM fermion and the SM-singlet NG boson $\varphi^{(0)}$. 
Hence, a characteristic signature of the process at the LHC  
   is a final state with two SM fermion jets and a missing energy from $\varphi^{(0)}$s.  
Note that this is analogous to a signature of superparticle pair production in the simplified supersymmetric models \cite{simplifiedSUSY}, 
   where a superparticle produced in pair decays to its partner SM particle and a stable neutralino. 
In order to obtain the current LHC constraints on KK-mode quarks, for example,  
   we may apply the ATLAS and CMS results from the search for squarks 
   with a process $pp \to {\tilde q} \bar{\tilde q}$, followed by ${\tilde q} \to q \tilde{\chi}_1^0$. 
A massless limit for neutralino corresponds to our case.
Although the production cross section for KK-mode quarks is a few times larger than that for squarks, 
   we roughly obtain  $m_{\rm KK} \gtrsim 2$ TeV for the KK-mode quarks from the current LHC results \cite{SUSY_LHC}.

%%%%%%%%%%%%%%%%%%%%%
\section{Conclusions and discussions}
%%%%%%%%%%%%%%%%%%%%%
\label{sec:6}

In Ref.~\cite{DWSM}, the authors of the present paper have proposed 
  a framework to construct the Domain-Wall SM which is defined in a non-compact 5D space-time.  
Considering localization mechanisms for the gauge field, 
  the Higgs field and its VEV, and the chiral fermion in the 5D flat Minkowski space,  
  we have derived the SM as the 4D effective theory at low energies. 
The model predicts the KK-modes for the SM particles, and we have briefly addressed 
  LHC phenomenology for KK-mode gauge bosons.

In this paper, we have investigated aspects of the Domain-Wall SM in detail. 
For concreteness, we have introduced two solvable examples to localize 
  all the SM particles in certain domains of the 5th dimension, 
  and have explicitly shown the KK-mode mass spectrum and eigenfunctions. 
One example predicts an infinite tower of the KK-mode of the SM gauge bosons,  
  while the number of KK-modes is finite in the other example.  
With explicit forms of the KK-mode eigenfunctions, we have derived the 4D effective Lagrangian 
  involving the KK-mode SM particles. 
The Domain-Wall SM offers a variety of interesting phenomenologies. 
Among others, we have addressed, in this paper, 
   the LHC phenomenology of the KK-mode gauge boson, 
  the effect of the KK-mode SM fermions on Higgs boson phenomenology, 
  and the KK-mode fermion search at the LHC with its decay 
  into a corresponding SM fermion and a NG boson 
  associated with a spontaneous breaking of the translational invariance in the 5th dimension.

In our solvable examples, we have introduced a special function $s(y)$ to localize the 5D gauge field, 
  as well as the 5D Higgs field and its VEV.   
In the theoretical point of view, we may seek a possible origin of $s(y)$.  
The second example of Eq.~(\ref{exp2}) is particularly interesting, 
  since it is expressed in terms of the kink solution when $m_V=m_\varphi$, 
\bea
  s(y)=  \frac{M}{\left[ \cosh(m_\varphi y) \right]^{2 \gamma}} 
   = M  \left(1- \frac{\sqrt{\lambda}}{m_\varphi} \, \varphi_{\rm kink}(y)^2 \right)^\gamma. 
\eea 
In the normal field theory sense, the parameter $\gamma$ is an integer. 
Since a KK-mode exists for $\gamma > 1$, we are interesting in the choice of $\gamma \geq 2$.  
Now we propose a unified picture of localizing all the SM fields by 
\bea 
   s(y) = M_G  \left(1- \frac{\sqrt{\lambda}}{m_\varphi} \, \varphi_{\rm kink}(y)^2 \right)^{\gamma_G}, \; \; \; \; 
   s_H(y) = M_H  \left(1- \frac{\sqrt{\lambda}}{m_\varphi} \, \varphi_{\rm kink}(y)^2 \right)^{\gamma_H}, 
\eea   
where $M_{G, H}$ are positive mass parameters, and $\gamma_{G, H}$ are integers.  
Note that this picture also introduces interactions between the physical modes in $\varphi$ and the gauge bosons. 
This phenomenology is worth considering.

In our analysis, we have implicitly assumed that all the SM fermions have the same domain-wall configuration. 
However, in general, SM chiral fermions can be localized around different points. 
Such a generalization opens up a possibility to solve the fermion mass hierarchy problem in the SM 
  from the wave-function overlapping, leading to an exponentially suppressed effective Yukawa coupling, 
  as proposed in Ref.~\cite{AS}.
Configurations of the domain-wall fermions reflect their effective gauge couplings with the KK-mode gauge bosons. 
Therefore, this ``geometry" relating to the fermion mass hierarchy can be tested at the future LHC experiment,  
  once a KK-mode gauge boson is discovered and its branching ratios into final state fermions are measured.

Since the graviton resides in the bulk, we also need to consider a localization of the graviton 
  to make the Domain-Wall SM phenomenologically viable. 
For this purpose, we may combine our model with the RS-2 scenario \cite{RS2} 
  with the Planck brane at $y=0$. 
Here we may identify the Planck brane as a domain-wall with the zero-width limit. 
The mass spectrum of the KK-modes of the SM fields is controlled 
  by the width of the domain-walls, and the current LHC results constrain it to be $\lesssim$(1 TeV)$^{-1}$. 
On the other hand, the width of 4D graviton is controlled by the AdS curvature $\kappa$ 
   in the RS-2 scenario and its experimental constraint is quite weak, $\kappa \gtrsim 10^{-3}$ eV \cite{RS2}. 
Therefore, we can take $\kappa \ll 1$ TeV and neglect the warped background geometry 
  in our setup of the Domain-Wall SM. 
The energy density from the SM domain-walls can affect the RS-2 background geometry. 
However, we expect this energy density is  of ${\cal O}(\Lambda^4)$ with $\Lambda={\cal O}$(1 TeV), 
  while the energy density of the Planck brane in the RS-2 scenario is given by ${\cal O}(M_P^2 \kappa^2)$ 
  with the reduced Planck mass of $M_P\simeq 2.4 \times10^{18}$ GeV.  
Therefore, we choose the AdS curvature in the range of $10^{-3}$ eV$\ll \kappa \ll$1 TeV 
  for the theoretical consistency of our scenario.

%%%%%%%%%%%%%%%%%%%%%%%%%%%%%%%%%%%%%%%%%
\section*{Acknowledgements}
%%%%%%%%%%%%%%%%%%%%%%%%%%%%%%%%%%%%%%%%%
The authors would like to thank Minoru Eto for addressing us a smart way to solve the KK mode equations. 
This work of N.O. is supported in part by the U.S. Department of Energy (DE-SC0012447).

%%%%%%%%%%%%%%

%%%%%%%%%%%%%%%%%%%%%%%%%%%%%%%%%%%%%%

\end{document}